\documentclass[12 pt,leqno]{article}%
\usepackage{latexsym}
\usepackage{amssymb,amsbsy,amsmath,amsfonts,amscd}
\usepackage{vmargin}
\usepackage{epsfig}
\usepackage{amsmath}
\usepackage{amsfonts}
\usepackage{amssymb}
\usepackage{graphicx}%
\providecommand{\U}[1]{\protect\rule{.1in}{.1in}}

\newtheorem{lemma}{Lemma}[section]

\newtheorem{theorem}{Theorem}[section]
\newtheorem{definition}{Definition}[section]
\newtheorem{proposition}{Proposition}[section]
\newtheorem{remark}{Remark}[section]

\newcommand{\N}{\ifmmode{{\rm I} \hskip -2pt {\rm N}}
    \else{\hbox{$I\hskip -2pt N$}}\fi}
\newcommand{\R}{\ifmmode{{\rm I} \hskip -2pt {\rm R}}
    \else{\hbox{$I\hskip -2pt R$}}\fi}

\newcommand{\Om}{\Omega}
\newcommand{ \vit}{\hbox{\bf u}}

\newcommand{ \vittest }{\hbox{\bf v}}
\newcommand{ \wit}{\hbox{\bf w}}

\newcommand{\g} {\nabla }

\newcommand{\x} {{\bf x}}
\newcommand{\BEQ} {\begin{equation} }
\newcommand{\EEQ} {\end{equation} }
\makeatletter
\@addtoreset{equation}{section} 
 
\begin{document}

\title{A high accuracy Leray-deconvolution model of turbulence and its limiting behavior.}
\author{William Layton\thanks{Department of Mathematics, University of Pittsburgh,
Pittsburgh, PA 15260, USA; wjl@pitt.edu, www.math.pitt.edu/\symbol{126}wjl,
partially supported by NSF grant DMS-0508260.}
\and Roger Lewandowski\thanks{IRMAR, UMR 6625,
Universit\'e Rennes 1,
Campus Beaulieu,
35042 Rennes cedex
FRANCE; 
Roger.Lewandowski@univ-rennes1.fr, 
http://perso.univ-rennes1.fr/roger.lewandowski/}}
\date{}
\maketitle

\begin{abstract}
In 1934 J. Leray proposed a regularization of the Navier-Stokes equations
whose limits were weak solutions of the NSE. Recently, a modification of the
Leray model, called the Leray-alpha model, has atracted study for turbulent
flow simulation. One common drawback of Leray type regularizations is their
low accuracy. Increasing the accuracy of a simulation based on a Leray
regularization requires cutting the averaging radius, i.e., remeshing and
resolving on finer meshes. This report analyzes a family of Leray type models
of arbitrarily high orders of accuracy for fixed averaging radius. We
establish the basic theory of the entire family including limiting behavior as
the averaging radius decreases to zero, (a simple extension of results known
for the Leray model). We also give a more technically interesting result on
the limit as the order of the models increases with fixed averaging radius.
Because of this property, increasing accuracy of the model is potentially
cheaper than decreasing the averaging radius (or meshwidth) and high order
models are doubly interesting.

\end{abstract}

MCS Classification : 76D05, 35Q30, 76F65, 76D03
\medskip


\section{Introduction}

In 1934 J. Leray \cite{Leray34a}, \cite{Leray34b} studied an interesting
regularization of the Navier-Stokes equations (NSE). He proved that the
\textit{regularized} NSE has a unique, smooth, strong solution and that as the
regularization length-scale $\delta\rightarrow0$, the regularized system's
solution converges (modulo a subsequence) to a\ weak solution of the
Navier-Stokes equations. This model has recently been attracting new interest
as continuum model upon which large eddy simulation can be based (see, e.g.,
the work of Geurts and Holm \cite{GH03} and Titi and co-workers \cite{CHOT05},
\cite{CTV05}, \cite{VTC05}). If $\overline{\wit}$ denotes a local spacial
average of the velocity $\mathbf{w}$ associated with filter length-scale
$\delta$\ , the classical Leray model is given by
\begin{equation}%
\begin{array}
[c]{l}%
\partial_{t}\wit+\overline{\wit}\cdot\nabla\wit-\nu\triangle\wit+\nabla
q=\mathbf{f},\\
\nabla\cdot\wit=0.
\end{array}
\label{Leray}%
\end{equation}
Leray chose $\overline{\wit}=g_{\delta}\star\wit$, where $g_{\delta}$ is a
Gaussian\footnote{By other choices of convolution kernel, differential
filters, sharp spectral cutoff and top-hat filters can be recovered.} with
averaging radius $\delta$. In $(\ref{Leray})$, $\nu$\ is the kinematic
viscosity, $(\wit,q)$\ denotes the model's velocity and pressure, $\Omega$\ is
the flow domain and $\mathbf{f}$\ is the body force, assumed herein to be
smooth and divergence free. We take $\Omega=(0,2\pi)^{3}$, the initial
condition
\[
\wit(0, \x)=\overline{\vit}_{0}(\x) \quad \text{  in } \quad \Omega\text{\ ,}%
\]
and impose periodic boundary conditions (with zero mean) on the solution (and
all problem data)
\[
\wit(t, \x+2\pi {\bf e}_{j})=\wit(t, \x)\quad \text{  and } \quad \int_{\Omega}\wit(t, \x)\, d\x=0.
\]

The Leray model is easy to solve using standard numerical methods for the
Navier-Stokes equations, e.g., \cite{LMNR06}, and, properly interpreted,
requires no extra or ad hoc boundary conditions in the non-periodic case.
However, as an LES model it has three main shortcomings:

\begin{itemize}
\item The Leray model's solution can suffer an accumulation of energy around
the cutoff frequency (Geurts and Holm \cite{GH03}). \ 
\end{itemize}

\begin{itemize}
\item The Gaussian filter is expensive to compute.
\end{itemize}

\begin{itemize}
\item The model $(\ref{Leray})$ has low accuracy on the smooth flow
components, e.g., $||\vit_{NSE}-\wit||=O(\delta^{2})$ , Section 4.
\end{itemize}

The accumulation of energy around the cutoff length-scale can be ameliorated
by new ideas in eddy viscosity which focus its effects on the smallest
resolved scales or by time relaxation models with similar motivations,
\cite{Guer}, \cite{SAK01a}, \cite{SAK01b}, \cite{SAK02}. The
expense of computing the filtered velocity is reduced (as proposed by Geurts
and Holm \cite{GH03}) if the Gaussian filter is replaced by a differential
filter (introduced into LES by Germano \cite{GermanoDiffFilters}). The
resulting combination of Leray model plus differential filter, called the
Leray-alpha model, has attracted an explosion of recent interest because of
its theoretical clarity and computational convenience. For recent work on the
Leray-alpha model, see Geurts and Holm \cite{GH03}, \cite{GH05}, Guermond and
Prudhomme \cite{GP05}, Cheskidov, Holm, Olson and Titi \cite{CHOT05}, Ilyin,
Lunasin and Titi \cite{ILT05}, Chepyzhov, Titi and Vishik \cite{CTV05},
\cite{VTC05} \ (among other works). Finally, there is the issue of the
accuracy of $(\ref{Leray})$ upon the large scales / smooth flow components
which must be improved if $(\ref{Leray})$ is to evolve from a descriptive
regularization to a predictive model.

In this paper we complement the above work on $(\ref{Leray})$ by presenting the
analysis of a new (and related) family of Leray-deconvolution models which
have arbitrarily high orders of accuracy and include the Leray-alpha model as
the zeroth order ($N=0$) case. To present the Leray-deconvolution models, some
preliminaries are needed. Although any reasonable filter can be used, for
definiteness, we fix the averaging to be by differential filter. Thus, given
$2\pi$-periodic free divergence field $\wit$ with zero mean value, its spacial
average over $O(\delta)$\ length-scales, denoted ${\overline{\wit}}$, is the
unique $2\pi$-periodic solution of the Stokes problem\footnote{This filter is
very close to the Gaussian filter; it is the first sub-diagonal Pad\'{e}
approximation thereof (as in the rational model \cite{GL00}); in the
Camassa-Holm/alpha model \cite{FHT01}, $(\ref{FIT07})$ is often called a
\textit{Helmholz filter}. One possible extension of this differential filter to
non-periodic boundary condition is given in Manica and Merdan \cite{MM06}.}
\begin{equation} \label{FIT07}
-\delta^{2}\triangle\overline{\wit}+\overline{\wit}+\nabla\pi=\wit\quad\text{
in }\mathbb{R}^{3},\quad\nabla\cdot\overline{\wit}=0,\quad\int_{\Omega
}\overline{\wit}=\mathbf{0}.
\end{equation}
It can be shown that $\pi$ is a constant in the equation above and therefore
the pressure term disappaers. Given a filter, the (\textit{ill-posed})
deconvolution problem is then:
\[
\text{given }\overline{\wit},\text{ find a (useful and stable)
\textit{approximation }to }\wit.
\]
Let $D_{N}$ denote the map: $\overline{\wit}\ \rightarrow$ \textit{chosen
approximation }to $\wit$, introduced in section 2. The approximate
de-convolution operators, $N=0,1,2,\cdot\cdot\cdot$ ,\ we consider have the
asymptotic accuracy and stability properties (see Section 2)%
\begin{gather}
\phi=D_{N}\overline{\phi}+O(\delta^{2N+2})\text{ , for smooth }\phi\text{,
}\label{EQ1}\\
||D_{N}||_{\mathcal{L}(\mathbf{H}_{0}\rightarrow\mathbf{H}_{0})}\leq
N+1<\infty,
\end{gather}
where
\begin{equation}
\mathbf{H}_{0}=\left\{  \mathbf{v}\in L_{\hbox{\footnotesize loc}}%
^{2}%
(\ifmmode{{\rm I} \hskip -2pt {\rm R}}\else{\hbox{$I\hskip -2pt R$}}\fi^{3}%
),\quad2\pi-\hbox{periodic},\quad\nabla\cdot\mathbf{v}=0,\quad\int_{\Omega
}\mathbf{v}=\mathbf{0}\right\}.
\end{equation}
The Leray-deconvolution model is then $\nabla\cdot\wit=0$\ and%
\begin{equation}
\partial_{t}\wit+(D_{N}\overline{\wit})\cdot\nabla\wit-\nu\triangle\wit+\nabla
q=\mathbf{f} \quad\text{ in } \quad \mathbb{R}^{3}\times(0,T), \quad
\int_{\Omega}\wit=\mathbf{0}, \label{Mod}%
\end{equation}
subject to initial and $2\pi-$periodic boundary conditions. Because of
$(\ref{EQ1})$, the formal accuracy of (1.4) on the smooth flow components as
an approximation of the Navier-Stokes equations is $O(\delta^{2N+2}%
),N=0,1,2,\cdot\cdot\cdot$, Section 4. When $N=0$, $D_{0}\overline
{\wit}=\overline{\wit}$\ so $(\ref{Mod})$ reduces to the Leray-alpha regularization.

The behavior of the model as $N$ increases for $\delta$\ fixed is beyond known
Leray-type theories and relevant for practical computation. Indeed, decreasing
$\delta$\ for fixed $N$\ requires reducing the computational meshwidth (a
process which increases the storage and computing time dramatically) and
resolving (1.1) or (1.2). On the other hand, increasing $N$\ for $\delta
$\ fixed requires only the solution of one additional Poisson or Stokes-type
problem per deconvolution step. The main theoretical contribution of this
paper is, in Section \ref{LDM}, to resolve this limiting behavior of the
model. We first prove existence and uniqueness of a smooth solution to $(\ref{Mod})$, then we show
that (modulo a subsequence), for fixed $\delta$, as $N\rightarrow\infty$ the
model solution $\wit=\wit(N)$ converges to a weak solution of the
Navier-Stokes equations. To our knowledge, this is the first result on the
limiting behavior of a family of turbulence models as the order of accuracy of
a family of models on the large scales increases. The difference between the
two limiting cases is, loosely speaking, that as $\delta\rightarrow0$\ ,
$\overline{\wit}\rightarrow\wit$ in every reasonable sense and the
deconvolution process inherits this: $D_{N}\overline{\wit}\rightarrow\wit$ for
$N$ fixed and as $\delta\rightarrow0$. However, since deconvolution is
ill-posed and the deconvolution operators $D_{N}$ are only an asymptotic (in
$\delta$\ for very smooth solutions and for $N$ fixed) inverse, the limiting
behavior as $N\rightarrow\infty$\ is more delicate.

Another new features include an $O(\delta^{2N+2})$ error bound for the energy
norm of the model's error (section \ref{ALDM}): $||\wit(\delta)-\vit_{NSE}||$,
provided the underlying solution of the Navier-Stokes equations is
sufficiently smooth, and an estimate of the time averaged error $<||\nabla
(\vit_{NSE}-\wit)||_{H}^{2}>^{\frac{1}{2}}$\ both for general weak solutions
of the Navier-Stokes equations and for weak solutions having the $k^{-\frac
{5}{3}}$ energy spectrum typical observed in fully developed turbulent flows.
(These first of the two estimates is connected to related results for
approximate deconvolution models in \cite{LL06a}, \cite{LL06b}, \cite{LL03}%
\ and \cite{DE06} and the second is an extension of the authors work in
\cite{LL06b}.)

Of course, ultimately, practical computations require analytic guidance in
balancing the computational meshwidth with $\delta$\ , $N$\ and other model
and algorithmic parameters.

\subsection{Notation and preliminaries}

We use $||\cdot||$ to denote the $L^{2}(\Omega)$\ norm and associated operator
norm. We always impose the zero mean condition $\int_{\Omega}\phi dx=0$ on
$\phi=\wit,p,\mathbf{f}$ and $\wit_{0}$. Recall that
\begin{equation}
\mathbf{H}_{0} = \left\{  \mathbf{w} \in L^{2}_{\hbox{\footnotesize loc}}
(\ifmmode{{\rm I} \hskip -2pt {\rm R}} \else{\hbox{$I\hskip -2pt R$}}\fi^{3}),
\quad2 \pi-\hbox{periodic}, \quad\nabla\cdot\mathbf{w} = 0, \quad\int_{\Omega
}\mathbf{w} = \mathbf{0} \right\}  .
\end{equation}
We also define the space function
\begin{equation}
\mathbf{H}_{1} = \left\{  \mathbf{w} \in H^{1}_{\hbox{\footnotesize loc}}
(\ifmmode{{\rm I} \hskip -2pt {\rm R}} \else{\hbox{$I\hskip -2pt R$}}\fi^{3}),
\quad2 \pi-\hbox{periodic}, \quad\nabla\cdot\mathbf{w} = 0, \quad\int_{\Omega
}\mathbf{w} = \mathbf{0} \right\}  .
\end{equation}
We shall define the space $\mathbf{H}_{s}$ for every $s$ in the same way. We
can thus expand the velocity in a Fourier series%
\[
\wit (\mathbf{x},t)=\sum_{\mathbf{k}}\widehat{\wit}(\mathbf{k}%
,t)e^{-i\mathbf{k. x}},\text{ where }\mathbf{k}\in\mathbb{Z}^{3}\text{ is the
wave-number.}%
\]
The Fourier coefficients are given by%
\[
\widehat{\wit}(\mathbf{k},t)=\frac{1}{(2\pi)^{3}}\int_{\Omega}%
\wit(x,t)e^{-i\mathbf{k . x}}d\mathbf{x.}%
\]
Magnitudes of $\mathbf{k}$ are defined by
\begin{align*}
|\mathbf{k}|  &  =\{|k_{1}|^{2}+|k_{2}|^{2}+|k_{3}|\}^{\frac{1}{2}},\\
|\mathbf{k}|_{\infty}  &  =\max\{|k_{1}|,|k_{2}|,|k_{3}|\}.
\end{align*}
The length-scale of the wave number $\mathbf{k}$\ is defined by
$\displaystyle l=\frac{2\pi}{|\mathbf{k}|_{\infty}}.$ Parseval's equality
implies that the energy in the flow can be decomposed by wave number as
follows. For $\wit\in\mathbf{H}_{0}$\ ,
\begin{gather*}
\frac{1}{(2\pi)^{3}}\int_{\Omega}\frac{1}{2}|\wit (x,t)|^{2}dx=\sum
_{\mathbf{k}}\frac{1}{2}|\widehat{\wit }(\mathbf{k},t)|^{2}=\\
=\sum_{k}\left(  \sum_{k-1<|\mathbf{k|\leq}k}\frac{1}{2}|\widehat
{\wit }(\mathbf{k},t)|^{2}\right)  ,\text{where }\mathbf{k}\in\mathbb{Z}%
^{3}\text{.}%
\end{gather*}
Moreover, for $s \in\ifmmode{{\rm I} \hskip -2pt {\rm R}}
\else{\hbox{$I\hskip -2pt R$}}\fi$,
\begin{equation}
\mathbf{H}_{s} = \left\{  \mathbf{v} \in H^{s}_{\hbox{\footnotesize loc}}
(\ifmmode{{\rm I} \hskip -2pt {\rm R}} \else{\hbox{$I\hskip -2pt R$}}\fi^{3}),
\quad2 \pi-\hbox{periodic}, \quad\nabla\cdot\mathbf{v} = 0, \quad\widehat\wit(
\mathbf{0} )= \mathbf{0}, \quad\sum_{\mathbf{k} } | \mathbf{k} |^{2s}
|\widehat{\wit }(\mathbf{k},t)|^{2} < \infty\right\}  .
\end{equation}
We define the $\mathbf{H}_{s}$ norms by
\begin{equation}
|| \wit ||^{2}_{s} = \sum_{\mathbf{k} } | \mathbf{k} |^{2s} |\widehat
{\wit }(\mathbf{k},t)|^{2},
\end{equation}
where of course $|| \wit ||^{2}_{0} = || \wit ||^{2}$. It can be shown that
when $s$ is an integer, $|| \wit ||^{2}_{s} = || \nabla^{s} \wit ||^{2}$ (see
\cite{DG95}).

\subsection{About the filter}

Let $\wit\in\mathbf{H}_{0}$ and $\overline{\wit}\in\mathbf{H}_{1}$ be the
unique solution to the Stokes problem.
\begin{equation}
-\delta^{2}\triangle\overline{\wit}+\overline{\wit}+\nabla\pi=\wit\quad\text{
in }\mathbb{R}^{3},\quad\nabla\cdot\overline{\wit}=0,\quad\int_{\Omega
}\overline{\wit}=\mathbf{0}. \label{filter}%
\end{equation}
It is usual in deconvolution studies to denote the filtering operation by $G$
so that $\overline{\mathbf{w}}=G\mathbf{w}$. Writing
\[
\wit(\mathbf{x},t)=\sum_{\mathbf{k}}\widehat{\wit}(\mathbf{k}%
,t)e^{-i\mathbf{k.x}},
\]
it is easily seen that $\nabla\pi=0$ and
\begin{equation}
\overline{\wit}(\mathbf{x},t)=\sum_{\mathbf{k}}{\frac{\widehat{\wit}%
(\mathbf{k},t)}{1+\delta^{2}|\mathbf{k}|^{2}}}e^{-i\mathbf{k.x}}%
\end{equation}
Then writing $\overline{\wit}=G(\wit)$, we see that in the corresponding
spaces of the type $\mathbf{H}_{s}$, the transfer function of $G$, denoted by
$\widehat{G}$ is the function
\[
\widehat{G}(\mathbf{k})={\frac{1}{1+\delta^{2}|\mathbf{k}|^{2}}},
\]
and we also can write on $\mathbf{H}_{s}$ spaces type
\begin{equation}
-\delta^{2}\triangle\overline{\wit}+\overline{\wit}=\wit\quad\text{ in
}\mathbb{R}^{3},\quad\nabla\cdot\overline{\wit}=0,\quad\int_{\Omega}%
\overline{\wit}=\mathbf{0}.
\end{equation}
Moreover, one notes that the transfer function depends only on the modulus of
the wave vector $\mathbf{k}$. Therefore, by noting $k=|\mathbf{k}|$, we shall
write in the following $\widehat{G}(k)$ instead of $\widehat{G}(\mathbf{k})$.

\section{\label{ADCO} Approximate de-convolution operators}

The de-convolution problem is central in image processing, \cite{BB98}. The
basic problem in approximate de-convolution is: given $\overline{\wit}$ solve
approximately for $\wit$:
\begin{equation}
G\wit=\overline{\wit}.
\end{equation}
Exact de-convolution is typically ill-posed. We consider the van Cittert
\cite{BB98} approximate de-convolution algorithm and associated operators,
introduced into LES modeling by Adams, Kleiser and Stolz, e.g., \cite{AS01},
\cite{AS02}, \cite{SA99}, \cite{SAK01a}, \cite{SAK01b}, \cite{SAK02}. For each
$N=0,1,\cdot\cdot\cdot,$ $\wit_{N}=D_{N}\overline{\wit}$ is computed as follows.

\begin{definition}
[van Cittert approximate de-convolution algorithm]Set $\wit_{0}=\overline
{\wit}$,

for $n=1,2,\cdot\cdot\cdot,N-1$,

perform $\wit_{n+1}=\wit_{n}+\{\overline{w}-G\wit_{n}\}$

Set $\wit_{N}=D_{N}\overline{\wit}$
\end{definition}

By eliminating the intermediate steps, we find%

\begin{equation}
D_{N}\wit=\sum_{n=0}^{N}(I-G)^{n}\wit.
\end{equation}
The approximate deconvolution operator $D_{N}$ is a \textit{bounded }operator
(Lemma \ref{STAB}) which approximates the \textit{unbounded} exact
deconvolution operator to high asymptotic accuracy $O(\delta^{2N+2})$\ on
subspaces of smooth functions (Lemma \ref{ERROR}). We begin by summarizing
from \cite{DE06}, \cite{BIL06}\ a few important, known properties of the
approximation deconvolution operator $D_{N}$ . \medskip

\begin{lemma}
\label{STAB} [Stability of approximate de-convolution] Let the averaging be
defined by $(\ref{FIT07})$ .\ Then $D_{N}$\ is a self-adjoint, positive semi-definite
operator on $\mathbf{H}_{0}$\ and
\[
||D_{N}||=N+1.
\]

\end{lemma}

\begin{lemma}
\label{ERROR} [Error in approximate de-convolution] Let $G$\ be given by the
differential filter $(\ref{FIT07})$. For any $\wit \in {\bf H}_{0},$%
\begin{align}
\wit-D_{N}\overline{\wit}  &  =(I-(-\delta^{2}\triangle+1)^{-1})^{N+1}\wit\\
&  =(-1)^{N+1}\delta^{2N+2}\triangle^{N+1}(-\delta^{2}\triangle+1)^{-(N+1)}%
\wit .\nonumber
\end{align}

\end{lemma}

\subsection{ Some calculations}

Since we consider the periodic problem, it is insightful visualize the
approximate de-convolution operators $D_{N}$\ in terms of the transfer
function of the operator $\widehat{D_{N}}$ . Since these are functions of
$\delta k$\ , where $k=|\mathbf{k}|$, and not $\delta$\ or $\mathbf{k}$, it is
appropriate to record them \textit{re-scaled by}
\[
k\leftarrow\delta k.
\]
Since $G=(-\delta^{2}\triangle+1)^{-1}$\ , we find, after rescaling,
\ $\displaystyle\widehat{G}(k)=\frac{1}{1+k^{2}}$. With that, the transfer
function of the first three deconvolution operators are%
\begin{align*}
\widehat{D_{0}}  &  =1,\\
\widehat{D_{1}}  &  =2-\frac{1}{k^{2}+1}=\frac{2k^{2}+1}{k^{2}+1},\text{
and}\\
\widehat{D_{2}}  &  =1+\frac{k^{2}}{k^{2}+1}+\left(  \frac{k^{2}}{k^{2}%
+1}\right)  ^{2}.
\end{align*}
These three are plotted together with the transfer function of exact
de-convolution ($k^{2}+1$) (in bold).

\begin{figure}[h]
\caption{{\protect\footnotesize Exact and approximate ($N=0,1,2)$)
de-convolution operators }}%
\label{profils}
\begin{center}
\includegraphics [scale=0.8]{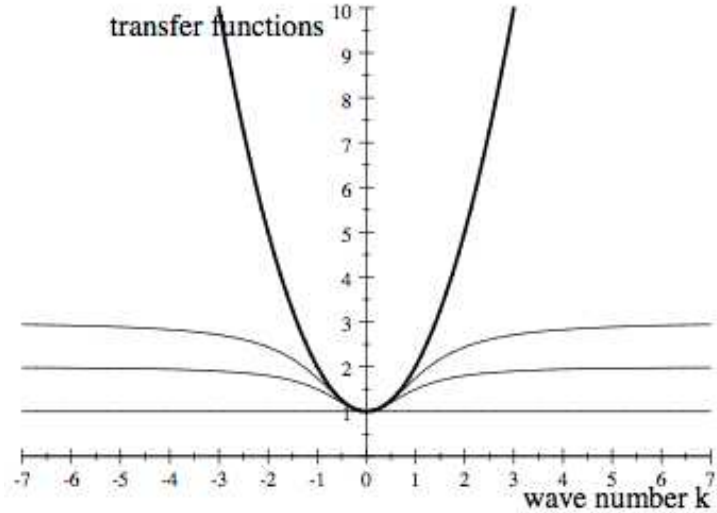}
\end{center}
\end{figure}

More generally, we find
\[
\widehat{D_{N}}(k):=\sum_{n=0}^{N} \left(  \frac{1}{1+k^{2}} \right)
^{n}=(1+k^{2}) \left[  1- \left(  \frac{k^{2}}{1+k^{2}} \right)  ^{N+1}
\right]  .
\]
\ The large scales are associated with the wave numbers near zero (i.e., $|k|
$ small). Thus, the fact that $D_{N}\ $is a very accurate solution of the
de-convolution problem for the large scales is reflected in the above graph in
that the transfer functions have high order contact near $k=0$. The key
observation in studying asymptotics of the model as $N\rightarrow\infty$\ is
that (loosely speaking) the region of high accuracy grows (slowly) as well as
$N$ increases.

The regularization in the nonlinearity involves a special combination of
averaging and deconvolution that we shall denote by $H_{N}$.

\begin{definition}
The truncation operator $H_{N}: \mathbf{H}_{0} \rightarrow\mathbf{H}_{0} $ is
defined by%
\[
H_{N}\wit:=D_{N}\overline{\wit }=(D_{N}\circ G)\wit .
\]

\end{definition}

\begin{proposition}
For each $N=0,1,\cdot\cdot\cdot$\ , the operator $H_{N}$\ is positive
semi-definite on $\mathbf{H}_{0}$\ . The operator $H_{N}:\mathbf{H}_{0}
\rightarrow\mathbf{H}_{0}$ is a compact operator. Moreover, $H_{N}$ maps
continuously $\mathbf{H}_{0}$ onto $\mathbf{H}_{2}$ . Further,%
\begin{align*}
H_{N}\wit  &  =\overline{D_{N}\wit },\\
||H_{N}||_{\mathcal{L} (\mathbf{H}_{0} \rightarrow\mathbf{H}_{0} ) }  &
=1\text{ , for each }N\geq0,\\
||H_{N}||_{\mathcal{L} (\mathbf{H}_{1} \rightarrow\mathbf{H}_{1})}  &  =1
\text{ , for each }N\geq0.
\end{align*}

\end{proposition}

\textbf{Proof} These properties are easily read off from the transfer function
$\widehat{H}_{N}(k)$\ of $H_{N}$\ which we give next. For example, compactness
follows since $|\widehat H_{N}(k)|\rightarrow0$ as $|k|\rightarrow\infty$.

\begin{remark}
Similarly, by the same proof, $H_{N}$ maps $\mathbf{H}_{s}$ into itself (see Lemma \ref{Lemme1} below), and
is compact. Since $\left\vert \widehat{H}_{N}(k)\right\vert \leq1$, one has
for each $s\geq0$
\begin{equation}
||H_{N}\wit||_{s}\leq||\wit||_{s}. \label{MER}%
\end{equation}

\end{remark}

The Fourier coefficients/transfer function of the operator $H_{N}$ are
similarly easily calculated \ to be (after rescaling by $k\leftarrow\delta
k$)
\begin{equation}
\widehat{H}_{N}(k)=1-\left(  \frac{k^{2}}{1+k^{2}}\right)  ^{N+1}.
\label{TRAN00}%
\end{equation}
They are plotted below for a few values of $N$.

\begin{figure}[h]
\caption{{\protect\footnotesize Transfer function of $H_{N}$, $N=0,10,50)$ }}%
\label{profils}
\begin{center}
\includegraphics [scale=0.8]{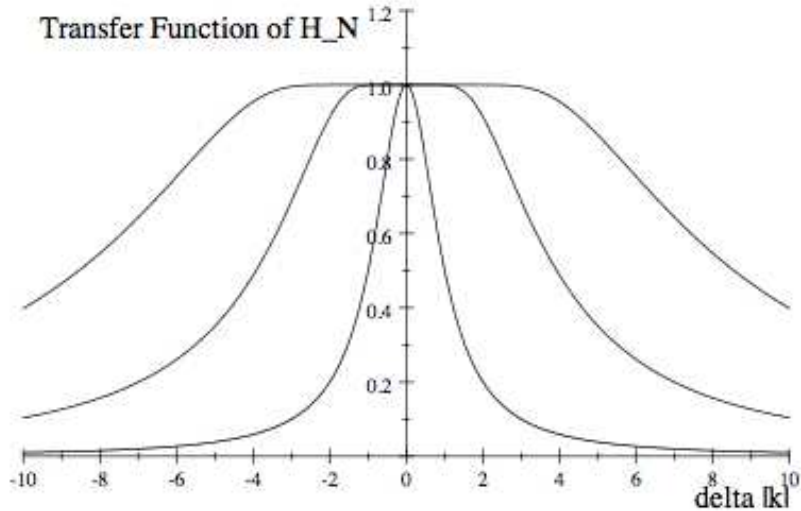}
\end{center}
\end{figure}

These plots are representative of the behavior of the whole family. Examining
the above graphs, we observe that $H_{N}(u)$ is very close to $u$ for the low
frequencies/largest solution scales and that $H_{N}(u)$ attenuates small
scales/high frequencies. The breakpoint between the low frequencies and high
frequencies is somewhat arbitrary. The following (from \cite{LN06a}) is
convenient for our purposes and fits our intuition of an approximate spectral
cutoff operator. We take for $k_{c}$ the frequency for which $\widehat{H}_{N}$
most closely attains the value $\frac{1}{2}.$

\begin{definition}
[Cutoff-Frequency]The cutoff frequency of $H_{N}$\ is
\[
k_{c}:=\text{greatest integer}\left(  \widehat{H}_{N}^{-1}\left(  \frac{1}{2}
\right)  \right)  .
\]
In other words, the frequency for which $\widehat{H}_{N}$ most closely attains
the value $\frac{1}{2}.$
\end{definition}

From the above explicit formulas, it is easy to verify that the cutoff
frequency grows to infinity slowly as $N\rightarrow\infty$ for fixed $\delta
$\ and as $\delta\rightarrow0$\ for fixed $N$. Other properties (whose proofs
are simple calculations) of the operator $H_{N}(\cdot)$\ follow similarly
easily from $\displaystyle \widehat{H}_{N}(k)=1- \left(  \frac{k^{2}}{1+k^{2}}
\right)  ^{N+1}$.

\begin{lemma}
For all $N\geq0,$ $\widehat{H}_{N}(k)$\ has the following properties:%
\begin{align*}
0  &  <\widehat{H}_{N}(k)\leq1,\\
|\widehat{H}_{N}(k)|  &  \rightarrow0\text{ as }|k|\rightarrow\infty,\\
\widehat{H}_{N}(0)  &  =1.
\end{align*}
Let $k_{c}$\ be cutoff frequency, then
\[
|k_{c}|\rightarrow\infty\text{ as }N\rightarrow\infty\text{ for fixed }%
\delta\text{ and as }\delta\rightarrow0\text{ for }N\text{ fixed.}%
\]
For any fixed value of $k$\ (or bounded set of values of $k$\ )%
\[
\widehat{H}_{N}(k)\rightarrow1\text{ as }N\rightarrow\infty\ \text{\ for
}\delta\ \text{fixed and as }\delta\rightarrow0\text{ \ for }N\ \text{fixed}%
\]
\ 
\end{lemma}

\textbf{Proof. } This follows from $(\ref{TRAN00})$. \medskip

To extract strong convergence of deconvolution, some restriction to the large
scales is needed (as shown clearly in the above two figures). One way to do
this is to consider the action of deconvolution operators on trigonometric polynomials.

\begin{definition}
The space \ $\mathbf{I} \! \mathbf{P}_{n}$ denotes the $2\pi$-periodic,
divergence-free, wit hzero mean value, vector, trigonometric polynomials of
degree $\leq n$ \ and $\Pi_{n}:\mathbf{H}_{0} \rightarrow\mathbf{I} \! \mathbf{P}_{n}$ \ is
the associated orthogonal projection.
\end{definition}

The model error is driven by the error in approximate deconvolution. With this
definition a critical convergence result on approximate deconvolution is possible.

\begin{proposition}
$H_{N}$ is a compact, symmetric positive semi-definite operator. For $n$
fixed, $H_{N}$\ is strictly positive definite on\ $\mathbf{I} \! \mathbf{P}_{n}$\ and
\[
H_{N}\rightarrow I\text{ on }\mathbf{I} \! \mathbf{P}_{n} \text{ in the strong operator norm
on }\mathbf{H}_{0}\text{.}%
\]
Given \ $n$ \ there is an $N_{0}(n)$ large enough such that%
\[
||\Pi_{n}\mathbf{v}||\leq2||H_{N}^{\frac{1}{2}}\mathbf{v} ||\text{ , for all
}N\geq N_{0}(n),
\]
Further,
\[
||\mathbf{v} ||\leq||H_{N}^{\frac{1}{2}}\mathbf{v}||\leq2||\mathbf{v} ||\text{
for all }\, \, \mathbf{v} \in P_{n}\text{.}%
\]

\end{proposition}

\textbf{Proof. } This follows from $(\ref{TRAN00})$. For example, Compactness
follows since $\widehat{H} _{N}(k)\rightarrow0$ as $k\rightarrow\infty.$

Other properties of the operator $H_{N}(\cdot)$\ follow similarly easily from
its transfer function.

\begin{proposition}
Let $\Pi_{k_{c}}$\ denote the orthogonal $\mathbf{H}_{0}$\ projection into

$span\{e^{i\mathbf{k\cdot x}}:|\mathbf{k}|\leq k_{c}\}.$ For all
$\wit \in\mathbf{H}_{0}$:
\begin{align}
(H_{N}\wit,\wit)  &  \geq C||\Pi_{k_{c}}\wit ||^{2},\\
(\wit -H_{N}\wit,\wit)  &  \geq C||(I-\Pi_{k_{c}})\wit||^{2}.\nonumber
\end{align}

\end{proposition}

\textbf{Proof} \ The claims follow from the definition of the cutoff
frequency, the explicit formula for the transfer function and direct calculation.

\bigskip The following properties are important to the analysis that follows.

\begin{lemma}
\label{OUF} Let $s \in\ifmmode{{\rm I} \hskip -2pt {\rm R}}
\else{\hbox{$I\hskip -2pt R$}}\fi$, $\wit  \in\mathbf{H}_{s}$. Then $H_{N}
(\wit) \in\mathbf{H}_{s+2}$.
\end{lemma}

\textbf{Proof}. Since $H_{N}(\wit)$ satisfies periodic boundary conditions, it
is divergence free and has zero mean value by construction. We check the
regularity. It is easy checked by using a Taylor expansion that there exists a
constant $C=C(\delta,N)$ such that
\begin{equation}
\forall\,\mathbf{k}\,\in\mathbb{Z}^{3},\,\mathbf{k}\not =0,\quad0\leq\hat
{H}_{N}(k)\leq{\frac{C(\delta,N)}{k^{2}}},\quad\hbox{with }k=\mathbf{k}.
\end{equation}
Therefore, for $\wit\in\mathbf{H}_{s}$, $\wit(\mathbf{x})=\sum_{\mathbf{k}%
\in\mathbb{Z}}\widehat{\wit}(\mathbf{k})e^{-i\mathbf{k\cdot x}}$
\begin{equation}
(\widehat{H}_{N}(k))^{2}k^{2(s+2)}|\wit(\mathbf{k})|^{2}\leq C(\delta
,N)^{2}k^{2s}|\wit(\mathbf{k})|^{2}. \label{241106}%
\end{equation}
Recall that here $\widehat{\wit}(\mathbf{0})=\mathbf{0}$. Therefore in
$(\ref{241106})$, $k\not =0$. Hence, by definition
\begin{equation}
||H_{N}(\wit)||_{\mathbf{H}_{s+2}}\leq C(\delta,N)||\wit||_{\mathbf{H}_{s}},
\label{BOOF}%
\end{equation}
and the proof is finished. \medskip

\begin{remark}
The constant $C(\delta,N)$ in the above proof blows up as $\delta\rightarrow0$
or as $N\rightarrow\infty$.
\end{remark}

\begin{lemma}
\label{Lemme1} 1) The operator $H_{N}$ maps continuously $\mathbf{H}_{s}$ into
$\mathbf{H}_{s}$ and
\begin{equation}
\label{NORM}|| H_{N}||_{\mathcal{L} (\mathbf{H}_{s},\mathbf{H}_{s}) } =1 .
\end{equation}

2) For every $\wit\in\mathbf{H}_{s}$, $H_{N}(\wit)$ converges strongly to
$\wit$ in $\mathbf{H}_{s}$ when $N\rightarrow\infty$.

3) $H_{N}$ commutes with the gradient operator: $\nabla H_{N}=H_{N}\nabla$.
\end{lemma}

\textbf{Proof. } 1) Let $\wit=(w^{1},w^{2},w^{3})\in\mathbf{H}_{s}$. One has
\[
||\wit||_{s}^{2}=\sum_{\mathbf{k}\in\mathbb{Z}^{3}}k^{2s}|\widehat
{w}(\mathbf{k})|^{2}.
\]
Let us note $\widehat{w}(\mathbf{k})=(w_{\mathbf{k}}^{1},w_{\mathbf{k}}%
^{2},w_{\mathbf{k}}^{3})$, $H_{N}(\wit)=(H_{N}(\wit)^{1},H_{N}(\wit)^{2}%
,H_{N}(\wit)^{3})$. Therefore, the $j^{\hbox {\footnotesize th}}$ component of
the vector field $H_{N}(\wit)$ is given by
\begin{equation}
H_{N}(\wit)^{j}=\sum_{\mathbf{k}\in\mathbb{Z}^{3}}\widehat{H}_{N}%
(k)w_{\mathbf{k}}^{j}e^{i\mathbf{k}.\mathbf{x}}.
\end{equation}
Consequently
\[
\nabla\cdot H_{N}(\wit)=\sum_{\mathbf{k}\in\mathbb{Z}^{3}}i\widehat{H}%
_{N}(k)k_{j}w_{\mathbf{k}}^{j}e^{i\mathbf{k}.\mathbf{x}}.
\]
When $\wit\in\mathbf{H}_{1}$, $\forall\,\mathbf{k}\in\mathbb{Z}^{3}$,
$k_{j}w_{\mathbf{k}}^{j}=0$. Then $\nabla\cdot H_{N}(\wit)=0$. Now recall
that
\[
\widehat{H}_{N}(k)=1-\left(  {\frac{k^{2}}{1+k^{2}}}\right)  ^{N+1}.
\]
One deduces that $|\hat{H}_{N}(k)|\leq1$ and $$k^{2s}|\hat{H}_{N}(k)\widehat
{w}(\mathbf{k})|^{2}\leq k^{2s}|\widehat{w}(\mathbf{k})|^{2}$$ yielding
$||H_{N}(\wit)||_{s}\leq||\wit||_{s}$. Hence 
$$H_{N}(\wit)\in\mathbf{H}_{s} \quad 
\hbox{and} \quad ||H_{N}(\wit)||_{\mathbf{H}_{s}}\leq||\wit||_{\mathbf{H}_{s}},$$ showing
$||H_{N}||_{\mathcal{L}(\mathbf{H}_{s},\mathbf{H}_{s})}\leq1$. Now let
$\mathbf{k}=(k_{1},k_{2},0)\in\mathbb{Z}$, $k_{1}\not =0$, $k=|\mathbf{k}|$
and define
\[
\displaystyle\widehat{\mathbf{w}}(\mathbf{k})={\frac{1}{k^{s}\sqrt
{1+(k_{2}/k_{1})^{2}}}}(1,-(k_{2}/k_{1}),0),\quad\wit_{\mathbf{k}}%
(\mathbf{x})=\widehat{\mathbf{w}}(\mathbf{k})e^{i\mathbf{k}.\mathbf{x}}.
\]
By construction $\wit_{\mathbf{k}}\in\mathbf{H}_{s}$, $||\wit_{\mathbf{k}%
}||_{\mathbf{H}_{s}}=1$, and one also has $||H_{N}\wit_{\mathbf{k}%
}||_{\mathbf{H}_{s}}=\widehat{H}_{N}(k)\rightarrow1$ as $k$ $\rightarrow
\infty$, making sure that $||H_{N}||_{\mathcal{L}(\mathbf{H}_{s}%
,\mathbf{H}_{s})}=1$. \smallskip

2) Let $\wit\in\mathbf{H}_{s}$. One has
\[
||\wit-H_{N}\wit||_{s}=\sum_{\mathbf{k}\in\mathbb{Z}^{3}}\left(  {\frac{k^{2}%
}{1+k^{2}}}\right)  ^{2(N+1)}k^{2s}|\widehat{\wit}(\mathbf{k})|^{2}%
\]
Let $\varepsilon>0$. Since $\wit\in\mathbf{H}_{s}$, there exists $K>0$ such
that
\[
\sum_{k\geq K}k^{2s}|\widehat{\wit}(\mathbf{k})|^{2}\leq{\frac{\varepsilon}%
{2}},
\]
yielding
\[
||\wit-H_{N}\wit||_{s}\leq\left(  {\frac{K^{2}}{1+K^{2}}}\right)
^{2(N+1)}||\wit||_{s}+{\frac{\varepsilon}{2}}%
\]
Now, as $K>0$, $K^{2}/(1+K^{2})<1$, there exists $N_{0}$ (which depends upon
$\wit$) such that $\forall\,N\geq N_{0}$,
\[
\left(  {\frac{K^{2}}{1+K^{2}}}\right)  ^{2(N+1)}||\wit||_{s}\leq
{\frac{\varepsilon}{2}}.
\]
Then $\forall\,N\geq N_{0}$, $||\wit-H_{N}\wit||_{\mathbf{H}_{s}}%
\leq\varepsilon$. This shows the strong $\mathbf{H}_{s}$ convergence of
$H_{N}\wit$ to $\wit$ as $N$ $\rightarrow\infty$.

3) The fact that $H_{N}$ commutes with $\nabla$ is obvious.

\begin{remark}
All the results in Lemma 2.5 above are uniform in $\delta$. Moreover, it easy
checked that for a fixed $N$, the same convergence result holds when $\delta$
goes to zero.
\end{remark}

\begin{remark}
\label{remcv} Let $\wit\in L^{2}([0,T],\mathbf{H}_{s})$. It easy checked by
using the same proof as in Lemma \ref{Lemme1} combined with the Lebesgue
monotone convergence theorem that $H_{N}(\wit)$ converges to $\wit$ in
$L^{2}([0,T],\mathbf{H}_{s})$ as $N$ goes to infinity.
\end{remark}

\section{\label{LDM} The Leray Deconvolution Model}

\subsection{Existence result}

The theory of Leray-deconvolution model begins, like the Leray theory of the
Navier-Stokes equations, with a clear global energy balance, and existence and
uniqueness of solutions.

Let $\vit_{0}\in\mathbf{H}_{0}$, $f\in\mathbf{H}_{-1}$. For $\delta>0$, let
the averaging be defined by $(\ref{filter})$. The problem we consider is the
following, for a fixed $T>0$, find $(\wit, q)$
\begin{equation}
\label{pb}\left\{
\begin{array}
[c]{l}%
\wit \in L^{2} ([0,T], \mathbf{H}_{1}) \cap L^{\infty}([0,T], \mathbf{H}_{0}),
\quad\partial_{t}\wit \in L^{2} ([0,T], \mathbf{H}_{-1})\\
q \in L^{2} ([0,T], L^{2}_{ {\hbox{\footnotesize per}, 0} }),\\
\partial_{t}\wit+ (H_{N} (\wit) \nabla) \wit-\nu\triangle\wit+\nabla
q=H_{N}(\mathbf{f}) \quad\text{ in } \mathbb{\mathcal{D}}^{\prime}( [0,T]
\times\ifmmode{{\rm I} \hskip -2pt {\rm R}}
\else{\hbox{$I\hskip -2pt R$}}\fi^{3}),\\
\wit ( \mathbf{x}, 0) = H_{N} (\vit_{0}) = \wit_{0}.
\end{array}
\right.
\end{equation}
where $L^{2}_{{\hbox{\footnotesize per}, 0} }$ denotes the scalar fields in
$L^{2}_{loc} (\ifmmode{{\rm I} \hskip -2pt {\rm R}}
\else{\hbox{$I\hskip -2pt R$}}\fi^{3})$, $2 \pi$-periodic with zero mean value.

\begin{theorem}
The problem $(\ref{pb})$ admits a unique solution $(\wit, q)$, where $\wit$
satisfies the energy equality
\begin{equation}
\label{ENER}\frac{1}{2}\left\Vert \wit (t)\right\Vert ^{2}+ \nu\int_{0}%
^{t}\int_{\Omega}|\nabla\wit|^{2}d\x dt^{\prime}=\frac{1}{2}\left\Vert H_{N}
(\vit_{0}\right\Vert ^{2}+\int_{0}^{t}\int_{\Omega} H_{N} (\mathbf{f}) .
\wit \, d\x dt^{\prime}.
\end{equation}
Moreover, $\wit \in L^{\infty}([0,T], \mathbf{H}_{1}) \cap L ^{2} ([0,T],
\mathbf{H}_{2})$
\end{theorem}

\textbf{Proof.} For simplification, one notes $||\cdot||_{p,B}$ the norm in
$L^{p}([0,T],B)$ and when $B=\mathbf{H}_{s}$ one shall note $||\cdot||_{p,s}$
for the simplicity. One also considers the space
\begin{equation}
\mathbf{V}=L^{2}([0,T],\mathbf{H}_{1})\cap L^{\infty}([0,T],\mathbf{H}%
_{0}),\quad||\wit||_{\mathbf{V}}=||\wit||_{2,1}+||\wit||_{\infty,0}.
\label{Espace}%
\end{equation}
Recall that (see \cite{RL06}),
\begin{equation}
\forall\wit\in\mathbf{V},\quad\forall\,r\in\lbrack2,6],\quad||\wit||_{{\frac
{4r}{3(r-2)}},L^{r}}\leq C||\wit||_{V}. \label{INTER}%
\end{equation}
We introduce the operators:
\begin{equation}
\begin{array}
[c]{l}%
\displaystyle A(\wit,\vittest)=\nu\int_{0}^{T}\int_{\Omega}\nabla
\wit.\nabla\vittest,\\
\displaystyle B_{N}(\wit,\vittest)=\int_{0}^{T}\int_{\Omega}(H_{N}%
(\wit)\nabla)\wit.\vittest=-\int_{0}^{T}\int_{\Omega}H_{N}(\wit)\otimes
\wit:\nabla\vittest.
\end{array}
\label{OP}%
\end{equation}
We first notice that
\begin{equation}
|A(\wit,\vittest)|\leq\nu||\wit||_{2,1}||\vittest||_{2,1}. \label{POUF}%
\end{equation}
Now assume that $\wit\in V$, $\vittest\in L^{2}([0,T],\mathbf{H}_{1})$. Using
lemma \ref{OUF} combined with an integration by parts and the Sobolev Theorem, it
is easy seen that there exits a constant $C$ which depends on $N$ and such
that one has
\begin{equation}
|B_{N}(\wit,\vittest)|\leq C||\wit||_{V}^{2}||\vittest||_{2,1}%
,\quad B_{N}(\wit,\wit)=0. \label{POUPOUF}%
\end{equation}
Here we have remarked that because $\mathbf{H}_{2}\subset(L^{\infty
}(\ifmmode{{\rm I} \hskip -2pt {\rm R}}\else{\hbox{$I\hskip -2pt R$}}\fi))^{3}%
$ (we are working in a 3D configuration), $\wit\in L^{\infty}([0,T],\mathbf{H}%
_{0})$ yields $H_{N}(\wit)\in L^{\infty}([0,T],\mathbf{H}_{2})$ (see
$(\ref{BOOF})$) and therefore,
\begin{equation}
H_{N}(\wit)\in L^{\infty}([0,T]\times
\ifmmode{{\rm I} \hskip -2pt {\rm R}}\else{\hbox{$I\hskip -2pt R$}}\fi),\quad
|| H_N(\wit) ||_{\infty,L^{\infty}}\leq C||\wit||_{V} \label{BOF2}%
\end{equation}
for a constant $C$ which depends on $N$. We now note
\begin{equation}
\mathbf{W}=\{\wit\in\mathbf{V},\,\,\partial_{t}\wit\in L^{2}([0,T],\mathbf{H}%
_{-1})\}.
\end{equation}
Notice that any $\wit\in\mathbf{W}$ is almost every-where equal to a function
in $C^{0}([0,T],\mathbf{H}_{0})$ (see in \cite{RT84}). Therefore, passing to a quotient space by keeping the same notations,  we have
\[
\mathbf{W}\subset C^{0}([0,T],\mathbf{H}_{0}).
\]
Finally when $\mathbf{f}\in L^{2}([0,T],\mathbf{H}_{-1})$, then
$H_{N}(\mathbf{f})\in L^{2}([0,T],\mathbf{H}_{1})$ and the variational
formulation of Problem $(\ref{pb})$ is:
\begin{equation}%
\begin{array}
[c]{l}%
\hbox{Find }\wit\in\mathbf{W}\quad\hbox{such that}\quad\wit(0,\mathbf{x}%
)=\wit_{0},\quad\hbox{and}\quad\forall\,\vittest\in L^{2}([0,T],\mathbf{H}%
_{1}),\\
\displaystyle\int_{0}^{T}<\partial_{t}\wit,\vittest>+B_{N}%
(\wit,\vittest)+A(\wit,\vittest)=\int_{0}^{T}\int_{\Omega}H_{N}(\mathbf{f}%
).\vittest\,d\x dt
\end{array}
\label{MEUH}%
\end{equation}
Since $(\ref{POUF})$ and $(\ref{POUPOUF})$ are satisfied, the existence of a
solution to Problem $(\ref{MEUH})$ can be derived thanks to the Galerkin
method. This is classical and the reader can look at \cite{LLe06} and
references inside or also in \cite{RT84} for a detailed descrition of such
kind of proof. By taking $\wit=\vittest$ as test vector field in
$(\ref{MEUH})$ (a legal operation here), one directly gets the energy estimate
$(\ref{ENER})$. Notice that thanks to the energy equality and $||H_{N}%
(\vit_{0})||\leq||\vit_{0}||$, $||H_{N}(\mathbf{f})||_{-1}\leq||\mathbf{f}%
||_{-1}$, $\wit$ does satisfy the following estimates:
\begin{equation}
||\wit||_{\infty,0}^{2}\leq||\wit_{0}||^{2}+{\frac{1}{\nu}}%
||\mathbf{f}||_{2,-1}^{2},\quad||\wit||_{2,1}^{2}\leq{\frac{1}{\nu}%
}||\wit_{0}||^{2}+{\frac{1}{\nu^{2}}}||\mathbf{f}||_{2,-1}^{2} \label{ESTIM}%
\end{equation}
in other words,
\begin{equation}
||\wit||_{V}\leq C(\nu,||\vit_{0}||,||\mathbf{f}||_{2,-1}).
\label{ESTIM2}%
\end{equation}
Here the bound  does not depend on $N$. Notice also that we can derive an
estimate for $\partial_{t}\wit$ in the space $L^{2}([0,T],\mathbf{H}_{-1})$.
However the bound for $\partial_{t}\wit$ in this space depends on $N$.

The pressure $q$ is recovered thanks the De Rham Theorem and its regularity
results from the fact that $\nabla q \in L^{2} ([0,T], \mathbf{H}_{-1})$.

We now check regularity. Let $D\wit$ and $Dq$ be a differential of $\wit$ and
$q$. Thanks to the periodic boundary conditions, one has
\begin{equation}
\partial_{t}D\wit+(H_{N}(\wit)\nabla)D\wit-\nu\Delta D\wit+\nabla
Dq=DH_{N}(\mathbf{f})-(DH_{N}(\wit)\nabla).\wit \label{DER}%
\end{equation}
One notes that $DH_{N}(\mathbf{f})\in L^{2}([0,T],(\mathbf{H}_{0})^{3})$,
$DH_{N}(\wit)\in L^{\infty}([0,T],(\mathbf{H}_{1})^{3})$. Since $\wit\in
L^{2}([0,T],\mathbf{H}_{0})$, by the Sobolev imbedding Theorem, $(DH_{N}%
(\wit)\nabla).\wit$ is periodic and in the space $L^{2}([0,T],(L^{3/2})^{9})$.
Since $1/6+2/3=5/6<1$, one has $$(DH_{N}(\wit)\nabla).\wit\in L^{2}%
([0,T],(L^{3/2})^{9})\subset L^{2}([0,T],\mathbf{H}_{-1}).$$ Then the equation
$(\ref{DER})$ admits a unique solution in the space ${\bf V}^3$. By using the same
technique as in \cite{LLe06}, it is easy to show that this solution is equal
to $D\wit$ showing that $\wit\in L^{\infty}([0,T],\mathbf{H}_{1})\cap
L^{2}([0,T],\mathbf{H}_{2})$. In particular, there is some constant $C$
(which depends on $N$) such that
\begin{equation}
||\wit||_{\infty,1}+||\wit||_{2,2}\leq C. \label{UTILE}%
\end{equation}

It remains to prove the uniqueness. Let $(\wit_{1}, q_{1})$ and $(\wit_{2},
q_{2})$ be two solutions, $\delta\wit = \wit_{1}-\wit_{2}$, $\delta q =
q_{2}-q_{1}$. Then one has
\begin{equation}
\partial_{t} \delta\wit + (H_{N} (\wit_{1}) \nabla) \delta\wit - \nu
\Delta\delta\wit + \nabla\delta q = (H_{N} (\delta\wit) \nabla) \wit_{2},
\end{equation}
and $\delta\wit = 0$ at initial time. All the terms in the equation above
being in $L^{2} ([0,T], \mathbf{H}_{-1})$, one can take $\delta\wit \in
\mathbf{W} \subset L^{2} ([0,T], \mathbf{H}_{1})$ as test,  a legal operation. Since $H_{N}
(\wit_1)$ is divergence free, one has
\[
\int_{0}^{T} \int_{\Omega}(H_{N} (\wit_{1}) \nabla) \delta\wit. \delta\wit =
0.
\]
Therefore, 
\BEQ \label{soir}  {\frac{d }{2dt}} \int_\Om | \delta\wit| ^{2} + \nu\int_{\Omega}| \nabla
\delta\wit |^{2} = \int_{\Omega}(H_{N} (\delta\wit) \nabla) \wit_{2}. \delta
\wit \EEQ
One has by a part integration,
$$\int_{\Omega}(H_{N} (\delta\wit) \nabla) \wit_{2}. \delta
\wit = -\int H_{N} (\delta\wit) \otimes \wit_{2} : \g \delta
\wit$$
By Young inequality, 
$$\int_{\Omega}(H_{N} (\delta\wit) \nabla) \wit_{2}. \delta
\wit = -\int H_{N} (\delta\wit) \otimes \wit_{2} : \g \delta
\wit \le {\nu \over 2} \int_\Om |\g \delta \wit |^2 + {1 \over 2 \nu} \int_\Om | H_N (\delta \wit) |^2 | \wit_2|^2. $$
Hence, using $(\ref{UTILE})$
\BEQ \label{soir2}  {\frac{d }{2dt}} \int_\Om | \delta\wit| ^{2} + {\nu \over 2} \int_{\Omega}| \nabla
\delta\wit |^{2} \le {1 \over 2 \nu} || \wit_2 ||_\infty^2 || H_N (\delta \wit ) ||^2 \EEQ
Therefore,
\[
{\frac{d }{2dt}} || \delta \wit ||^2 
 \le C (t) || \delta\wit|| ^{2} ,
\]
where $\displaystyle C(t) = {1 \over 2 \nu} || \wit_2 ||_\infty^2 \in L^1 ([0,T])$. 
We conclude that $\delta\wit = 0$ thanks to Gronwall's Lemma.

\begin{remark}
It easy checked that the solution $(\wit, q)$ is such that $\wit \in C^{0}
([0,T], \mathbf{H}_{1})$.
\end{remark}

\begin{remark}
By iterating the process obove, it is easy to prove that when $\mathbf{f} \in
C^{\infty}([0,T] \times\ifmmode{{\rm I} \hskip -2pt {\rm R}}
\else{\hbox{$I\hskip -2pt R$}}\fi)$ and $\vit_{0} \in C^{\infty}%
(\ifmmode{{\rm I} \hskip -2pt {\rm R}} \else{\hbox{$I\hskip -2pt R$}}\fi)$
(both space periodic), then $\wit \in C^{\infty}([0,T] \times
\ifmmode{{\rm I} \hskip -2pt {\rm R}} \else{\hbox{$I\hskip -2pt R$}}\fi)$.
\end{remark}

\begin{remark}
The Galerkin approximations to the solution $\wit$ are under the form
\[
\wit_{n}=\sum_{|k|\leq n}\widehat{\wit}_{n}(t,\mathbf{k})e^{i\mathbf{k}%
.\mathbf{x}},\quad\wit_{n}(0,\mathbf{x})=\Pi_{n}(\wit_{0})(\mathbf{x}).
\]
where the vector $({\wit}_{n}(t,-\mathbf{n}),...,{\wit}_{n}(t,\mathbf{n}))$ is
a solution of an ODE and is of class $C^{1}$ on $[0,T]$ according to the
Cauchy Lipchitz Theorem. We know on one hand that the sequence $(\wit_{n}%
)_{n\in\ifmmode{{\rm I} \hskip -2pt {\rm N}}\else{\hbox{$I\hskip -2pt N$}}\fi}%
$ converges strongly to $\wit$ in $L^{2}([0,T),\mathbf{H}_{0})$ whether the
sequence $(\partial_{t}\wit_{n})_{n\in
\ifmmode{{\rm I} \hskip -2pt {\rm N}}\else{\hbox{$I\hskip -2pt N$}}\fi}$
converges weakly to $\partial_{t}\wit$ in $L^{2}([0,T),\mathbf{H}_{-1})$ . On
the other hand, for any smooth periodic field $\boldsymbol{\phi}$ with
$\nabla\cdot\boldsymbol{\phi}=0$ and $\boldsymbol{\phi}(T,\mathbf{x}%
)=\mathbf{0}$, since $\wit_{n}\in C^{1}([0,T],\mathbf{I}\!\mathbf{P}_{n})$, a
legal part integration yields
\[
\displaystyle%
\begin{array}
[c]{l}%
\displaystyle\int_{0}^{T}<\partial_{t}\wit_{n},\boldsymbol{\phi}>=\int_{0}%
^{T}\int_{\Omega}\partial_{t}\wit_{n}(t,\mathbf{x}).\,\boldsymbol{\phi
}(t,\mathbf{x})\,d\mathbf{x}dt=\\
\displaystyle\int_{\Omega}\Pi_{n}(\wit_{0})(\mathbf{x}).\,\boldsymbol{\phi
}(0,\mathbf{x})\,d\mathbf{x}-\int_{0}^{T}\int_{\Omega}\wit_{n}(t,\mathbf{x}%
).\,\partial_{t}\boldsymbol{\phi}(t,\mathbf{x})\,d\mathbf{x}dt.
\end{array}
\]
Passing to the limit when $n$ $\rightarrow\infty$ yields
\begin{equation}
\displaystyle\int_{0}^{T}<\partial_{t}\wit,\boldsymbol{\phi}>=\int_{\Omega
}\wit_{0}(\mathbf{x}).\,\boldsymbol{\phi}(0,\mathbf{x})\,d\mathbf{x}-\int
_{0}^{T}\int_{\Omega}\wit(t,\mathbf{x}).\,\partial_{t}\boldsymbol{\phi
}(t,\mathbf{x})\,d\mathbf{x}dt.  \label{RRT}%
\end{equation}

\end{remark}

\subsection{Limiting behavior of the Leray-deconvolution model}

The models considered are intended as approximations to the Navier-Stokes
equations. Thus the limiting behavior of the models solution is of primary
interest. There are two natural limits: $\delta\rightarrow0$\ and
$N\rightarrow\infty$. The first is the normal analytic question of turbulence
modeling and considered already by J. Leray in 1934. The question of the
behavior of the model's solution as $N\rightarrow\infty$\ is much more
unclear, however.

In practical computations, cutting $\delta$\ means re-meshing and increasing
the memory and run time requirements greatly while increasing $N$\ simply
means solving one more shifted Poisson problem per deconvolution step. Thus,
increasing the accuracy of the model is much easier than decreasing its
resolution. On the other hand, the van Cittert deconvolution procedure itself
is an asymptotic approximation rather than a convergent one and has a large
error at smaller length scales.

Before developing these results some preliminary definitions are needed.

The notion of weak solution is due to J. Leray \cite{Leray34a} who called them
\textit{turbulent solutions}. Recall that the Navier-Stokes equations are the
following:
\begin{equation}%
\begin{array}
[c]{l}%
\partial_{t}\vit+\vit\cdot\nabla\vit-\nu\triangle\vit+\nabla p=\mathbf{f},\\
\nabla\cdot\vit=0,\\
\vit(0,\mathbf{x})=\vit_{0}%
\end{array}
\label{Leray1}%
\end{equation}

\begin{definition}
\label{WSL} [Weak solutions of Navier-Stokes Equations] Let $\vit_{0}%
\in\mathbf{H}_{0}$ $\mathbf{f} \in L^{2}([0,T], \mathbf{H}_{-1} )$. A
measurable vector field $\vit(t, \mathbf{x}):\Omega\times\lbrack
0,T]\rightarrow\mathbf{R}^{3}$ is a \textsl{weak solution} to the
Navier-Stokes equations if

(i) $\vit \in\mathbf{V}$ (where $\mathbf{V}$ is defined by $(\ref{Espace})$),
\medskip

(ii) $\vit$ satisfies the integral relation:%
\begin{equation}
\label{FVF}%
\begin{array}
[c]{l}%
\displaystyle - \int_{0}^{T}\int_{\Omega}\vit \, . \, \partial_{t}
\boldsymbol{\phi} +\nu\int_{0}^{T} \int_{\Omega}\nabla\vit :\nabla
\boldsymbol{\phi} + \int_{0}^{T} \int_{\Omega}(\vit \nabla) \vit \, . \,
\boldsymbol{\phi}\\
\hskip 6.5cm \displaystyle =\int_{0}^{T}<{\bf f},\boldsymbol{\phi} >\ dt)-
\int_{\Omega}\vit_{0}\, . \boldsymbol{\phi}(0, \cdot)
\end{array}
\end{equation}
\ for all $\boldsymbol{\phi} \in(C^{\infty}([0,T] \times
\ifmmode{{\rm I} \hskip -2pt {\rm R}} \else{\hbox{$I\hskip -2pt R$}}\fi))^{3}%
$, space periodic with $\boldsymbol{\phi} (T, \mathbf{x}) = 0$ forall
$\mathbf{x} \in\ifmmode{{\rm I} \hskip -2pt {\rm R}}
\else{\hbox{$I\hskip -2pt R$}}\fi^{3}$ and such that $\nabla\cdot
\boldsymbol{\phi} = 0$.

\medskip

(iii) [Leray's inequality/the energy inequality] for any $t\in\lbrack0,T]$
\begin{equation}
\label{ENIN}\frac{1}{2}||\vit(t, \cdot)||^{2}+\nu\int_{0}^{t}||\nabla
\vit(t^{\prime}, \cdot)||^{2}dt^{\prime}\leq\frac{1}{2}||\vit_{0}||^{2}+
\int_{0}^{t} <\mathbf{f} , \vit > dt^{\prime},
\end{equation}

(iv) $\lim\limits_{t\rightarrow0}\ ||\vit (t, \cdot)-\vit_{0}||=0$.
\end{definition}

A $N$ being given, we denote by $(\wit_{N}, q_{N})$ the unique solution to the
Leray-Deconvolution problem $(\ref{pb})$, $\wit_{N} (0, \mathbf{x}) =
\wit_{0,N} (\mathbf{x}) = H_{N} (\vit_{0})(\mathbf{x})$. We prove the following

\begin{theorem}
There exists a sequence $(N_{j})_{j \in\ifmmode{{\rm I} \hskip -2pt {\rm N}}
\else{\hbox{$I\hskip -2pt N$}}\fi}$ be such that $(\wit_{N_{j}})_{j
\in\ifmmode{{\rm I} \hskip -2pt {\rm N}} \else{\hbox{$I\hskip -2pt N$}}\fi}$
converges to a weak solution $\vit$ to the Navier-Stokes Equations. The
convergence is weak in $L^{2}([0,T], \mathbf{H}_{1})$ and strong in
$L^{2}([0,T], \mathbf{H}_{0})$
\end{theorem}

\textbf{Proof.} Thanks to the bound $(\ref{ESTIM})$, from the sequence
$(\wit_{N})_{N\in
\ifmmode{{\rm I} \hskip -2pt {\rm N}}\else{\hbox{$I\hskip -2pt N$}}\fi}$ one
can extract a subsequence $(\wit_{N_{j}})_{j\in
\ifmmode{{\rm I} \hskip -2pt {\rm N}}\else{\hbox{$I\hskip -2pt N$}}\fi}$ which
converges to some $\vit$, weakly in $L^{2}([0,T],\mathbf{H}_{1})$. In the
following, we shall denote by $(\wit_{j})_{j\in
\ifmmode{{\rm I} \hskip -2pt {\rm N}}\else{\hbox{$I\hskip -2pt N$}}\fi}$ this
subsequence and we have to show that $\vit$ is a weak solution to the
Navier-Stokes equations as defined in Definition $(\ref{WSL})$. \medskip

Thanks to $(\ref{INTER})$ combined with $(\ref{ESTIM})$, the sequence
$(\wit_{j})_{j \in\ifmmode{{\rm I} \hskip -2pt {\rm N}}
\else{\hbox{$I\hskip -2pt N$}}\fi}$ is also bounded in $L^{8/3} ([0,T],
(L^{4})^{3} )$, while $(H_{N_{j}}(\wit_{j}))_{j \in
\ifmmode{{\rm I} \hskip -2pt {\rm N}} \else{\hbox{$I\hskip -2pt N$}}\fi}$ is
bounded in $L^{\infty} ([0,T], \mathbf{H}_{0} )$. Hence the sequence
$(H_{N_{j}}(\wit_{j}) \otimes\wit_{j})_{j \in
\ifmmode{{\rm I} \hskip -2pt {\rm N}} \else{\hbox{$I\hskip -2pt N$}}\fi}$ is
bounded in $L^{8/3} ([0,T], (L^{4/3})^{9} )$ making $B_{N_{j}} (\wit_{j},
\cdot)$ bounded in the space $L^{8/5} ([0,T], \mathbf{H}_{-2})$. \medskip

Obviously,
$(A (\wit_{j}, \cdot))_{j \in\ifmmode{{\rm I} \hskip -2pt {\rm N}}
\else{\hbox{$I\hskip -2pt N$}}\fi}$ is bounded in $L^{2} ([0,T],
\mathbf{H}_{-1})$ as well as $(H_{N_{j}} ( { \bf f} ) ) _{j \in \N} $
Writing the equation for $\wit_{j}$ under the form,
\begin{equation}
\label{EQZ}\partial_{t} \wit_{j} = - A (\wit_{j}, \cdot) - B_{N_{j}}
(\wit_{j}, \cdot) + H_{N_{j}} (\mathbf{f),}%
\end{equation}
one deduces that the sequence $(\partial_{t} \wit_{j})_{j \in
\ifmmode{{\rm I} \hskip -2pt {\rm N}} \else{\hbox{$I\hskip -2pt N$}}\fi}$ is
bounded in $L^{8/5} ([0,T], \mathbf{H}_{-2})$. Since one has $\mathbf{H}_{1}
\subset\mathbf{H}_{0} \subset\mathbf{H}_{-2}$, the first injection being
continuous compact and dense, the second being continuous and dense, one
deduces from Aubin-Lions Lemma (see in \cite{JS87}) that the sequence
$(\wit_{j})_{j \in\ifmmode{{\rm I} \hskip -2pt {\rm N}}
\else{\hbox{$I\hskip -2pt N$}}\fi}$ is compact in $L^{8/3} ([0,T],
\mathbf{H}_{0})$, and by very classical arguments in all $L^{p} ([0,T],
\mathbf{H}_{0})$, $p<\infty$ and also in $L^{q} ([0,T], (L^{4})^{3} )$ for
$q<8/3$. \medskip

Even extracting an other subsequence still denoted by $(\wit_{j})_{j
\in\ifmmode{{\rm I} \hskip -2pt {\rm N}} \else{\hbox{$I\hskip -2pt N$}}\fi}$,
the sequence $(\wit_{j})_{j \in\ifmmode{{\rm I} \hskip -2pt {\rm N}}
\else{\hbox{$I\hskip -2pt N$}}\fi}$ converges to $\vit$ almost everywhere (use
the Lebesgue inverse Theorem). Notice that thanks to Fatou's Lemma, for almost
every $t \in[0,T]$, one has
\begin{equation}
\label{IINE13}||\vit (t, \cdot)||\le\liminf|| \wit_{j}
(t, \cdot)||.
\end{equation}
Then by $(\ref{ESTIM})$, $\vit \in\mathbf{V}$.
\bigskip

We have to show that $\vit$ satisfies points (i) and (ii) of Definition
\ref{WSL} to prove that it is a weak solution to the Navier-Stokes equations,
knowing already that it satisfies point (i). \medskip

We check point (ii). Let $\boldsymbol{\phi} \in C^{\infty}([0,T]
\times\ifmmode{{\rm I} \hskip -2pt {\rm R}}
\else{\hbox{$I\hskip -2pt R$}}\fi)$, space periodic with $\boldsymbol{\phi}
(T, \mathbf{x}) = 0$ for all $\mathbf{x} \in
\ifmmode{{\rm I} \hskip -2pt {\rm R}} \else{\hbox{$I\hskip -2pt R$}}\fi^{3}$
and $\nabla\cdot\boldsymbol{\phi} = 0$ (we denote by $\mathbf{T}$ the space
made of such fields). Notice that $\boldsymbol{\phi} \in L^{2}([0,T],
\mathbf{H}_{1})$ and therefore can be used as test vector field in formulation
$(\ref{MEUH})$. It is obvious that one has
\begin{equation}
\label{Lim1}\lim_{j \rightarrow\infty}A (\wit_{j}, \boldsymbol{\phi}) =
A(\vit, \boldsymbol{\phi}) = \int_{0}^{T} \int_{\Omega}\nabla\vit (t,
\mathbf{x}) : \nabla\boldsymbol{\phi} (t, \mathbf{x}) \, d\mathbf{x} dt.
\end{equation}
By using the result in Remark \ref{remcv}, it is also obvious that
\begin{equation}
\label{Lim2}\lim_{j \rightarrow\infty} \int_{0}^{T} \int_{\Omega}H_{N}
(\mathbf{f}) (t, \mathbf{x}) \, . \, \wit_{j} (t, \mathbf{x})\, d\mathbf{x} dt
= \int_{0}^{T} < \mathbf{f}, \vit >,
\end{equation}
as well as thanks to $(\ref{RRT})$ and since $(H_{N} (\vit_{0}))_{N
\in\ifmmode{{\rm I} \hskip -2pt {\rm N}} \else{\hbox{$I\hskip -2pt N$}}\fi}$
converges to $\vit_{0}$ in the space $\mathbf{H}_{0}$ (Lemma \ref{Lemme1}),
\begin{equation}
\label{Lim3}\lim_{j \rightarrow\infty} \int_0^T <\wit_{j}, \boldsymbol{\phi} > =
\int_{\Omega}\vit_{0} (\mathbf{x}) \, . \, \boldsymbol{\phi} (0, \mathbf{x})
\, d\mathbf{x} - \int_{0}^{T} \int_{\Omega}\vit (t, \mathbf{x}) \, . \,
\partial_{t} \boldsymbol{\phi} (t, \mathbf{x}) \, d\mathbf{x} dt.
\end{equation}
It remains to pass to the limit in the non linearity. Notice that
\[
|| H_{N_{j}} (\wit_{j}) - \vit ||_{2,0} \le|| H_{N_{j}} (\wit_{j} - \vit)
||_{2,0} + || H_{N_{j}} (\vit) - \vit ||_{2,0} \le|| \wit_{j} - \vit ||_{2,0}
+ || H_{N_{j}} (\vit) - \vit ||_{2,0}.
\]
Hence it is easy deduced that $(H_{N_{j}} (\wit_{j}))_{j \in
\ifmmode{{\rm I} \hskip -2pt {\rm N}} \else{\hbox{$I\hskip -2pt N$}}\fi}$
converges towards $\vit$ in $L^{2}([0,T], \mathbf{H}_{0})$ and consequently
$(H_{N_{j}} (\wit_{j}) \otimes\wit_{j})_{j \in
\ifmmode{{\rm I} \hskip -2pt {\rm N}} \else{\hbox{$I\hskip -2pt N$}}\fi}$
converges towards $\vit \otimes\vit$ in $L^{1} ([0,T], (L^{1})^{9})$.
Therefore,
\begin{equation}
\label{Lim4}\lim_{j \rightarrow\infty} B(\wit_{j}, \boldsymbol{\phi}) =
B(\vit, \boldsymbol{\phi}) = -\int_{0}^{T} \int_{\Omega}\vit \otimes\vit :
\nabla\boldsymbol{\phi} = \int_{0}^{T} \int_{\Omega}(\vit \nabla) \vit \, . \,
\boldsymbol{\phi}.
\end{equation}
Combining $(\ref{Lim1})$, $(\ref{Lim2})$, $(\ref{Lim3})$ and $(\ref{Lim4})$
makes sure that $(\ref{FVF})$ is satisfied.

\medskip We now check point (iii). We already know that $(\partial
_{t}\wit)_{j\in
\ifmmode{{\rm I} \hskip -2pt {\rm N}}\else{\hbox{$I\hskip -2pt N$}}\fi}$ is
bounded in the space $L^{8/3}([0,T],\mathbf{H}_{-2})$. Thus, up to a
subsequence, it converges weakly in this space to some $g$. Passing to the
limit in $(\ref{RRT})$ for $\boldsymbol{\phi}\in\mathbf{T}$ , one sees that
$g$ satisfies
\begin{equation}
\displaystyle\int_{0}^{T}<g,\boldsymbol{\phi}>=\int_{\Omega}\vit_{0}%
(\mathbf{x}).\,\boldsymbol{\phi}(0,\mathbf{x})\,d\mathbf{x}-\int_{0}^{T}%
\int_{\Omega}\vit(t,\mathbf{x}).\,\partial_{t}\boldsymbol{\phi}(t,\mathbf{x}%
)\,d\mathbf{x}dt \label{RRT1}%
\end{equation}
Hence the following relation holds in the space $L^{8/3}([0,T],\mathbf{H}%
_{-2})$ for all $t\in\lbrack0,T]$:
\begin{equation}
\vit(t,\cdot)=\vit_{0}(t,\cdot)+\int_{0}^{t}g(s)ds
\end{equation}
Consequently, $\vit\in C^{0}([0,T],\mathbf{H}_{-2})$. Since $\mathbf{H}%
_{0}\subset\mathbf{H}_{-2}$, the injection being dense, and $\vit\in
L^{\infty}([0,T],\mathbf{H}_{0})$, $\vit(t,\cdot)$ does exists for each
$t\in\lbrack0,T]$ and is weakly continuous from $[0,T]$ into $\mathbf{H}_{0}$.
Moreover, one has $\vit(0,\cdot)=\vit_{0}$. Now by weak convergence of
$(\wit_{j})_{j\in
\ifmmode{{\rm I} \hskip -2pt {\rm N}}\else{\hbox{$I\hskip -2pt N$}}\fi}$ to
$\vit$ in $L^{2}([0,T],\mathbf{H}_{1})$, one has
\begin{equation}
\int_{0}^{t}\int_{\Omega}|\nabla\vit|^{2}\leq\liminf_{j\rightarrow\infty}%
\int_{0}^{t}\int_{\Omega}|\nabla\wit_{j}|^{2}. \label{HIT}%
\end{equation}
It is easy checked that $(H_{N_{j}}(\mathbf{f}.\wit_{j})_{j\in
\ifmmode{{\rm I} \hskip -2pt {\rm N}}\else{\hbox{$I\hskip -2pt N$}}\fi}$
converges to $\mathbf{f})$, $\vit$ $\in$ $L^{1}([0,T]\times\Omega)$, and we
have previously shown that $(H_{N_{j}}(\vit_{0}))_{j\in
\ifmmode{{\rm I} \hskip -2pt {\rm N}}\else{\hbox{$I\hskip -2pt N$}}\fi}$
converges towards $\vit_{0}$. Since each $\wit_{j}$ satisfies the energy
equality $(\ref{ENER})$, $(\ref{IINE13})$ combined with $(\ref{HIT})$ 
ensures that the energy inequality $(\ref{ENIN})$ holds for almost every $t$ in
$[0,T]$. We have to prove that it holds for every $t\in\lbrack0,T]$. Let
$t\in\lbrack0,T]$ and $(t_{k})_{k\in
\ifmmode{{\rm I} \hskip -2pt {\rm N}}\else{\hbox{$I\hskip -2pt N$}}\fi}$ be a
sequence that converges to $t$ and that satisfies $(\ref{ENIN})$ for each $k$.
We already know that $(\vit({t_{k}},\cdot))_{k\in
\ifmmode{{\rm I} \hskip -2pt {\rm N}}\else{\hbox{$I\hskip -2pt N$}}\fi}$
converges weakly to $\vit(t,\cdot)$ in the space $\mathbf{H}_{0}$. Therefore,
\[
||\vit(t,\cdot)||\leq\liminf_{k\rightarrow\infty}||\vit(t_{k},\cdot)||.
\]
One deduces from this fact that $(\ref{ENIN})$ is satisfied because
\[
s\rightarrow\int_{0}^{s}||\nabla\vit(s^{\prime},\cdot)||^{2}\,ds^{\prime
},\quad s\rightarrow\int_{0}^{s}<\mathbf{f}(s^{\prime}),\vit(s^{\prime}%
,\cdot)>ds^{\prime}.
\]
are continous functions of $s$.

\medskip We finish by checking point (iv). We already knows that $\vit(t,
\cdot)$ converges weakly to $\vit_{0}$ in $\mathbf{H}_{0}$. Therefore
\[
|| \vit_{0} || \le\liminf_{t \rightarrow0} || \vit (t, \cdot) ||.
\]
As $|| \nabla\vit (t, \cdot) ||^{2}$ and $<{\bf f}, \vit>$ are both in
$L^{1}([0,T])$, one deduces from the energy inequality $(\ref{ENIN})$ that
\[
\limsup_{t \rightarrow0} || \vit (t, \cdot) || \le|| \vit_{0} ||
\]
Therefore, $\displaystyle \lim_{t \rightarrow0} || \vit (t, \cdot) || = ||
\vit_{0} ||$ which combined to the weak convergence in $\mathbf{H}_{0}$
garanties the strong convergence, in particular $\displaystyle \lim_{t
\rightarrow0} || \vit (t, \cdot) - \vit_{0} || = 0$ and the proof is finished.

\begin{remark}
By using same arguments as above, one can show that for a fixed $N$, the
sequence of solution to $(\ref{pb})$ converges to a weak solution of the
Navier-Stokes solution when $\delta$ goes to zero. The proof is left to the reader.
\end{remark}

\section{\label{ALDM} Accuracy of Leray and Leray-deconvolution models}

The accuracy of a regularization model as $\delta\rightarrow0$ is typically
studied in two ways. The first, called a posteriori analysis in turbulence
model validation, is to obtain via direct numerical simulation (or from a DNS
database) a "truth" solution of the Navier-Stokes equations, then, to solve
the model numerically for varying values of $\delta$\ and compute directly
various modeling errors, such as $\vit-\wit$ and $\overline{\vit}-\wit$ . The
second approach, known as \'{a} priori analysis in turbulence model validation
studies (and is exactly an experimental estimation of a model's consistency
error), is to compute the residual of the true solution of the Navier-Stokes
equations (obtained from a DNS database) in the model. For example, to assess
the consistency error of the Leray (and Leray-alpha model) model, the
Navier-Stokes equations is rewritten to make the Leray model appear on the LHS
as%
\[
\partial_{t}\vit+\overline{\vit}\cdot\nabla\vit-\nu\triangle\vit+\nabla
p-\mathbf{f}=\nabla\cdot\lbrack\overline{\vit}\vit-\vit\vit]\quad\text{ in
}[0,T]\times\Omega.
\]
The Leray-model's consistency error tensor is then $\boldsymbol{\tau}%
_{Leray}(\vit,\vit):=\overline{\vit}\vit-\vit\vit$ . Analysis of the modeling
error in various deconvolution models, various norms and diverse settings in
\cite{LL06a}, \cite{LL06b}, \cite{BIL06} and \cite{DE06} has shown that the
energy norm of the model error, $||\overline{\vit}_{NSE}-\wit_{model}||$ or
$||\vit_{NSE}-\wit_{model}|||$\ as appropriate, is driven by the consistency
error tensor $\boldsymbol{\tau}$\ rather than $\nabla\cdot\boldsymbol{\tau}$.
\medskip

Thus, an analysis of a model's consistency error analysis evaluates
$||\overline{\vit}\vit-\vit \vit||.$ In the analysis of consistency errors,
there are three interesting and important cases. Naturally, the case where
$\vit$\ is a general, weak solution of the Navier-Stokes equations is most
interesting and equally naturally nothing can be expected within current
mathematical techniques beyond very weak convergence to zero, possibly modulo
a subsequence. Next is the case of smooth solutions (the classical case for
evaluating consistency errors analytically). The case of smooth solutions is
important for transitional flows and regions in non-homogeneous turbulence and
it is an important analytical check that the LES model is very close to the
Navier-Stokes equations on the large scales. The third case (introduced in
\cite{LL06b}) is to study time averaged consistency errors using the
intermediate regularity observed in typical time averaged turbulent velocities.

For the Gaussian filter it is known (e.g., Chapter 1 in \cite{BIL06}) that for
smooth $\phi$ , $\phi-(g_{\delta}\star\phi)=O(\delta^{2})$\ so that the Leray
model's consistency error is second order accurate in $\delta$ on the smooth
velocity components: $\boldsymbol{\tau}_{Leray}=(\overline{\vit}%
-\vit)\vit=O(\delta^{2})$. \ This simple calculation shows that the
consistency error is dominated by the error in the regularization of the
convecting velocity. Thus, improving the accuracy of a Leray-type
regularization model hinges on improving the accuracy of the regularization.
For the differential filter, from (1.2), $\phi-\overline{\phi}=\delta
^{2}(-\triangle\overline{\phi})$ so the consistency error of the Leray-alpha
model (Geurts and Holm \cite{GH03}) is also $O(\delta^{2})$. To make \ this
more precise, we begin with a simple lemma of \cite{LL06a}, given in the case
of scalar fields for the simplicity, the same result being true in the case of
smooth vector fields. (We include a short proof of this lemma for completeness.)

\begin{lemma}
Let $\overline{u}=(-\delta^{2}\triangle+1)^{-1}u $. Then for any
derivative $\partial^{|\beta|}u/\partial x^{\beta}$\ , with multi-index
$|\beta|\geq0$%
\begin{align*}
||\frac{\partial^{|\beta|}}{\partial x^{\beta}}(u-\overline{u})||  &
=\delta^{2}||\triangle\frac{\partial^{|\beta|}}{\partial x^{\beta}}%
\overline{u}||\text{ }\leq\delta^{2}||\triangle\frac{\partial^{|\beta|}%
}{\partial x^{\beta}}u||\text{ , and }\\
||\frac{\partial^{|\beta|}}{\partial x^{\beta}}(u-\overline{u})||  &
\leq\frac{\delta}{2}||\nabla\frac{\partial^{|\beta|}}{\partial x^{\beta}}u||.
\end{align*}

\end{lemma}

\textbf{Proof} The first follows from the definition of averaging equation,
stability of averaging and the fact that derivatives commute under periodic
boundary conditions. For the second, note that the error equation satisfies
the equation%
\[
-\delta^{2}\triangle(u-\overline{u})+(u-\overline{u})=-\delta^{2}\triangle
u\text{ ,}%
\]
which is an identity. Differentiating through this equation shows that the
derivatives of the error also satisfy the same equation (with the derivative
of $u$ on the RHS instead of $u$ ). Multiplying by $(u-\overline{u})$,
integrating over $\Omega$, integrating by parts and using the Cauchy Schwarz
inequality gives%
\begin{align*}
\delta^{2}||\nabla(u-\overline{u})||^{2}+||(u-\overline{u})||^{2}  &
\leq\delta^{2}||\nabla u||||\nabla(u-\overline{u})||\\
&  \leq\delta^{2}||\nabla(u-\overline{u})||^{2}+\frac{\delta^{2}}{4}||\nabla
u||^{2},
\end{align*}
and the result follows for the the error $(u-\overline{u})$. Since the
derivatives of the error satisfy the same equations as the error, the same
proof shows works for the derivatives of the error as well.

\subsection{Consistency error of the Leray and Leray-alpha model}

The Leray /Leray-alpha model is the case $N=0$\ in the family of
Leray-deconvolution models so we shall denote the consistency error tensor of
the Leray-alpha model by $\tau_{0}(\vit,\vit)$ where
\begin{align*}
\tau_{0}(\vit,\vit)  &  :=\overline{\vit} \otimes \vit-\vit \otimes \vit \text{ , }\\
\text{where }\overline{\vit}  &  =(-\delta^{2}\triangle+1)^{-1}\vit \text{ (for
Leray-alpha),}\\
\text{and }\overline{\vit}  &  =g_{\delta}\star \vit \text{ (for Leray).}%
\end{align*}
For the simplicity, we shall write $\overline{\vit} \otimes \vit-\vit \otimes \vit  = 
\overline{\vit}\vit-\vit  \vit$. 
 
Using estimates of both filter's accuracy, a sharp estimate of the consistency
error of both can be given.

\begin{proposition}
The consistency error of the Leray model and Leray-alpha model satisfy, for
smooth $u$,
\[
\text{ }\int_{\Omega}|\tau_{0}(\vit,\vit)|d\x \leq C\delta||\vit ||\min\{||\nabla
\vit ||,\delta||\triangle \vit||\}.
\]

\end{proposition}

\textbf{Proof} By the Cauchy-Schwarz inequality%
\[
\int_{\Omega}|\tau_{0}(\vit,\vit)|d\x \leq||\vit-\overline{\vit}||||\vit||.
\]
Both filters satisfy
\[
||\vit-\overline{\vit}||\leq C\delta\min\{||\nabla \vit||,\delta||\triangle \vit||\},
\]
e.g., for the Leray-alpha model $C=1$, from which the result follows.\ \qquad

Naturally, for a general weak solution of the Navier-Stokes equations, the
indicated norms on the above RHS may or may not be finite at specific times.
Time averaged values, however, are always well defined and can be estimated in
terms of model parameters and the Reynolds number as in \cite{LL06b}.

\begin{definition}
Let $<\cdot>$\ denote long time averaging, given by%
\begin{equation} \displaystyle 
<\phi>:={\rm LIM}_{ T \rightarrow \infty} \frac{1}{T}\int_{0}^{T}\phi(t)dt, 
\end{equation}
where {\rm LIM} denotes the generalized limit for bounded functions introduced in \cite{FMRT01} 

\end{definition}

\begin{definition} 

Let \ $\varepsilon(\vit)$ and $\varepsilon$\ denote respectively the energy
dissipation rate of the flow (the unknown true solution of the Navier-Stokes
equations) and its time averaged value, defined by%
\begin{align*}
\varepsilon(\vit)(t)  &  =\frac{1}{L^{3}}\int_{\Omega}\nu|\nabla \vit|^{2}dx\text{ ,
and }\\
\varepsilon &  =<\varepsilon(\vit)(t)>.
\end{align*}

\end{definition}

\begin{lemma}
If ${\bf f}\in L^{\infty}([0,T], {\bf H}_{-1})$ then $<\varepsilon(\vit)><\infty$ and
$<\varepsilon(\vit)><\infty$.
\end{lemma}

\textbf{Proof} The NSE result is standard, e.g., Doering and Gibbon
\cite{DG95}. The result for the model is proven the same way (divide the
energy equality by $T$\ and take the limit as $T\rightarrow\infty$. Using
Gronwall's inequality, $\frac{1}{T}||\vit (T)||^{2}\rightarrow0$. The remainder
follows from the Cauchy-Schwarz inequality).

Using the estimate of the regularization's accuracy, a sharp estimate of the
Leray model's consistency error can be given. Naturally, for a general weak
solution of the Navier-Stokes equations, the indicated norms on the RHS may or
may not be finite at specific times.

\begin{proposition}
The time averaged consistency error of the Leray model and Leray-alpha model
satisfy, for smooth $\vit$,
\[
\text{ }<\int_{\Omega}|\tau_{0}(\vit,\vit)|d\x >\leq C\delta<||\vit||^{2}>^{\frac{1}{2}%
}\min\{<||\nabla \vit||^{2}>^{\frac{1}{2}},\delta<||\triangle \vit||^{2}>^{\frac
{1}{2}}\}.
\]

\end{proposition}

\textbf{Proof} By the Cauchy-Schwarz inequality%
\[
\int_{\Omega}|\tau_{0}(\vit,\vit)|d\x\leq||\vit-\overline{\vit}||||\vit||.
\]
Both filters satisfy
\[
||\vit-\overline{\vit}||\leq C\delta\min\{||\nabla \vit||,\delta||\triangle \vit||\},
\]
e.g., for the Leray-alpha model $C=1$. Integrate this in time, apply the
temporal Cauchy-Schwarz inequality, take limits superior and the result follows since the generalized limit is below the limit superior.

It is useful to estimate the RHS of these bounds in terms of the Reynolds
number and (after time averaging) for a general weak solution of the
Navier-Stokes equations. To do so a selection of the reference velocity must
be made. In this generality the natural choice is%
\[ 
U:=<\frac{1}{L^{3}}\int_{\Omega}|\vit(t , \x )|^{2}d\x >^{\frac{1}{2}}.
\]
With this choice we have
\begin{align*}
&  <\frac{1}{L^{3}}\int_{\Omega}|\tau_{0}(\vit,\vit)|d\x >\leq C\delta<\frac
{1}{L^{\frac{3}{2}}}||\nabla \vit||\frac{1}{L^{\frac{3}{2}}}||\vit||>\text{ , or}\\
&  <\frac{1}{L^{3}}\int_{\Omega}|\tau_{0}(\vit,\vit)|d\x>\leq C\frac{\delta}%
{\sqrt{\nu}}<\frac{\nu}{L^{3}}||\nabla \vit||^{2}>^{\frac{1}{2}}<\frac{1}{L^{3}%
}||\vit||^{2}>^{\frac{1}{2}}\text{.}%
\end{align*}
Rewriting these in terms of the non-dimensionalized quantities gives
\[
<\frac{1}{L^{3}}\int_{\Omega}|\tau_{0}(\vit,\vit)|d\x >\leq\frac{\delta}{\sqrt{\nu}%
}\varepsilon^{\frac{1}{2}}U^{\frac{1}{2}}.
\]
Now, in turbulent flow typically $\varepsilon\approx\frac{U^{3}}{L}$\ (e.g.,
by dimensional analysis, Pope \cite{Pope}, and experiment, \cite{S84},
\cite{S98}. The estimate $\varepsilon\leq C\frac{U^{3}}{L}$ has also ben
proven directly from the Navier-Stokes equations, e.g., \cite{CKG01},
\cite{CD92}, \cite{DF02}, \cite{Wang97}). Thus we have
\[
<\frac{1}{L^{3}}\int_{\Omega}|\tau_{0}(\vit,\vit)|d\x>\leq C\frac{\delta}{\sqrt{\nu}%
} \left (\frac{U^{3}}{L} \right )^{\frac{1}{2}}U^{\frac{1}{2}}=C\frac{\delta}{L}Re^{\frac
{1}{2}}U^{\frac{3}{2}};
\]
Since both LHS and RHS are quadratic in $U$\ the most incisive form of this
estimate is%
\[
<\frac{1}{U^{2}L^{3}}\int_{\Omega}|\tau_{0}(\vit,\vit)|d\x >\simeq\frac{\delta}%
{L}\frac{Re^{\frac{1}{2}}}{U^{\frac{1}{2}}}.
\]
Related and more detailed estimates can be obtained in the case of
homogeneous, isotropic turbulence the techniques introduced in \cite{LL06b}.

\begin{remark}
The above presentation has been based on considering the Leray
regularization's solution as an approximation of the true (unfiltered)
solution of the Navier-Stokes equations. If the filter is at least formally
invertible, Geurts and Holm \cite{GH03}, \cite{GH05} have shown that by a
change of variables a related Leray-regularized problem can be constructed
whose solution is an approximation to the average velocity $\overline{u}$.
This change of variables alters the definition of the consistency error as
follows. If $\overline{u}=A^{-1}u$ then $u=A\overline{u}$\ . Since $w\simeq
u$\ ( as $\delta\rightarrow0$ ) $\overline{w}=A^{-1}w\simeq\overline{u}$\ and
the equation $\overline{w}$\ is a Leray-regularization approximating the
filtered solution of the Navier-Stokes equations. Changing variables by
$\wit =AA^{-1}\wit =A\overline{\wit}$\ in the model gives (after simplification):%
\[
\overline{\wit}_{t}+\nabla\cdot(\overline{\overline{\wit }A\overline{\wit}}%
)-\nu\triangle\overline{\wit }+\nabla\overline{q}=\overline{ \bf f} \quad \text{in }%
\Omega.
\]
To calculate \textit{this} model's consistency error, rewrite the space
filtered Navier-Stokes equations as
\[
\overline{\vit}_{t}+\nabla\cdot(\overline{\overline{\vit}A\overline{u}}%
)-\nu\triangle\overline{\vit}+\nabla\overline{p}-\overline{\bf f}=\nabla\cdot
\lbrack\overline{\overline{\vit}A\overline{\vit}-\vit \vit}]\text{ .}%
\]
In this way we define the tensor%
\[
\overline{\tau_{0}(\vit,\vit)}:=\overline{\overline{u}A\overline{\vit}-\vit \vit }\text{ }%
\]
as the consistency error of the Leray and Leray-alpha model considered as an
approximation to $\overline{u}$ . However, this change of variables is an
isomorphism and thus does not alter the final result on the model's
consistency error.
\end{remark}

\subsection{Consistency error of the Leray-deconvolution model}

\bigskip We now consider the consistency error of the Leray-deconvolution
model and show that the (asymptotic as $\delta\rightarrow0$ ) consistency
error is $O(\delta^{2N+2})$. To identify the consistency error tensor, the
Navier-Stokes equations is rearranged%
\[
\vit_{t}+D_{N}(\overline{\vit})\cdot\nabla \vit-\nu\triangle \vit+\nabla p-{\bf f}=\nabla
\cdot\lbrack D_{N}(\overline{\vit})\vit-\vit \vit ] \quad \text{in }\Omega\times(0,T). %
\]
The LHS is the Leray-deconvolution model and the RHS is the \textit{residual
of the true solution of the Navier-Stokes equations in the model}. Thus, the
consistency error tensor is%
\[
\tau_{N}(\vit,\vit):=D_{N}(\overline{\vit})\vit-\vit \vit \quad \text{for }N=0,1,\cdot\cdot
\cdot\text{\ .}%
\]
As in the Leray model, adapting the analysis in \cite{LL06a}, \cite{LL06b} to
the present case, the model error is driven by the model's consistency error
$\tau_{N}(\vit,\vit)$ rather than $\nabla\cdot\tau_{N}(\vit,\vit)$. Since $\tau_{N}%
=(D_{N}(\overline{\vit})-\vit)\vit$ the consistency error of the Leray-deconvolution
model is dominated by the deconvolution error. As before, there are three
cases: a general weak solution, solutions with the regularity typically
observed in homogeneous, isotropic turbulence (the deconvolution error is
estimated in \cite{LL06b} and the same estimates hold here) and, to assess
accuracy on the large scales, very smooth solutions. In this case the
deconvolution error is bounded in \cite{DE06}, \cite{LL06b} and induces a high
order consistency error bound, given next.

\begin{proposition}
The consistency error of the $N^{th}$ Leray-deconvolution model is
$O(\delta^{2N+2})$ ; it satisfies%
\begin{align*}
\int_{\Omega}|\tau_{N}(\vit,\vit)|d\x \text{ }  &  \leq\delta^{2N+2}||\triangle
^{N+1}(-\delta^{2}\triangle+1)^{-(N+1)}\vit ||||\vit ||,\\
&  \leq C\delta^{2N+2}||\triangle^{N+1}\vit ||||\vit ||.
\end{align*}

\end{proposition}

\textbf{Proof} By the Cauchy-Schwarz inequality and Lemma 2.3%
\begin{align*}
\int_{\Omega}|\tau_{N}(\vit,\vit)|dx  &  \leq||\vit-D_{N}(\overline{\vit})||||\vit||\\
&  \leq\delta^{2N+2}||\triangle^{N+1}(-\delta^{2}\triangle+1)^{-(N+1)}%
\vit||||\vit||\text{ .}%
\end{align*}

In the case of homogeneous, isotropic turbulence, precise estimates of time
averaged consistency errors can be given following \cite{LL06b}.

\section{Conclusions}

The Leray-regularization approach shows theoretical promise and has appealing
simplicity. The tests of the Leray regularization in Geurts and Holm
\cite{GH03} have also been positive and the initial tests in \cite{LMNR06} the
higher order models $(\ref{pb})$ have shown deamatic improvements over the Leray
model $(\ref{Leray})$. Extensive and systematic computational testing is the natural
next step. However, there are also substantial theoretical questions open as
well. The first is to develop a similarity theory for the models (paralleling
the similarity theory in \cite{Mus96} for the Smagorinsky model, in
\cite{FHT01} for the alpha model and in \cite{LN06a}, \cite{LN06b} for other
deconvlution models) to better understand how the regularization in the model
truncates solution scales. The second natural question is to study other
filters and deconvolution operators. At this point, we believe that extension
to many (but not all) other filters should present only technical problems.
Extension to other deconvolution operators is important. Due to the many
approaches to solution of ill-posed problems, this extension will most likely
be done on a case by case basis. In practical computations, most likely both
$\delta$\ and $N$\ will be varying. Thus, a detailed understanding of limiting
behavior in both variables is also an important open problem.

\bigskip

\textbf{Acknowledgement}. The solution to the accuracy problem of Leray models
studied herein, the family of Leray-deconvolution models, is an ingenious idea
of A. Dunca as is the observation that the Leray-theory of existence,
uniqueness and convergence as $\delta\rightarrow0$ extends to the new family.

R. Lewandowski thanks the Math Department of Pittsburgh University for the
warm and reapeted welcome.

\end{document}